\documentclass[11pt]{article}
\usepackage{amssymb}
\usepackage{graphicx}
\usepackage{amsfonts}
\usepackage{amsmath}
\usepackage{longtable,lscape}
\usepackage{amscd}
\usepackage{mathrsfs}
\usepackage{latexsym}
\usepackage{bbm}
\usepackage{float}
\usepackage{url}

\newcommand{\captionfonts}{\footnotesize}
\makeatletter  
\long\def\@makecaption#1#2{%
  \vskip\abovecaptionskip
  \sbox\@tempboxa{{\captionfonts #1: #2}}%
  \ifdim \wd\@tempboxa >\hsize
    {\captionfonts #1: #2\par}
  \else
    \hbox to\hsize{\hfil\box\@tempboxa\hfil}%
  \fi
  \vskip\belowcaptionskip}
\makeatother 

\title{{\bf Explaining versus Describing Human Decisions} \\ {\bf Hilbert Space Structures in Decision Theory}}
\author{Sandro Sozzo\\ School of Business and Centre IQSCS \\ University Road LE1 7RH \\ 
Leicester (United Kingdom) \\ Email address: \url{ss831@le.ac.uk}}

\date{}

\begin{document}

\maketitle

\begin{abstract}
\noindent
Despite the impressive success of quantum structures to model long-standing human judgement and decision puzzles, the {\it quantum cognition research programme} still faces challenges about its explanatory power. Indeed, quantum models introduce new parameters, which may fit empirical data without necessarily explaining them. Also, one wonders whether more general non-classical structures are better equipped to model cognitive phenomena. In this paper, we provide a {\it realistic-operational foundation of decision processes} using a known decision-making puzzle, the {\it Ellsberg paradox}, as a case study. Then, we elaborate a novel representation of the Ellsberg decision situation applying standard quantum correspondence rules which map realistic-operational entities into quantum mathematical terms. This result opens the way towards an independent, foundational rather than phenomenological, motivation for a general use of quantum Hilbert space structures in human cognition.
\end{abstract}

\section{Introduction\label{intro}}
Traditional cognitive theories systematically apply classical set-theoretic structures to model human judgements and decisions under uncertainty. This is particularly evident in theories of rational decision-making, like expected utility theory, where Bayesian, or Kolmogorovian \cite{k1933}, models of probability directly follow from axioms on agents' preferences \cite{vnm1944,s1954}.

However, several {\it cognitive puzzles} have been  discovered in empirical tests, which provide evidence of systematic deviations from Kolmogorovian probability structures (see, e.g., \cite{bb2012}). For example, Kahneman and Tversky identified a {\it conjunction fallacy} in human probability judgements, namely, the law of monotonicity of Kolmogorovian probability does not generally hold in this kind of judgements \cite{kst1982}. Also, in human decision-making, Tversky and Shafir proved that the law of total Kolmogorovian probability does not hold in the {\it disjunction effect} \cite{kt2000}, while Allais and Ellsberg indicated that people do not always choose by maximizing an expected utility with respect to a Kolmogorovian probability measure \cite{e1961}.

As a consequence of the puzzles above, traditional theories using Kolmogorovian structures, though normatively compelling, are descriptively flawed, which led several authors to elaborate alternative proposals able to more efficiently and realistically represent human behaviour. This was the starting point of the {\it bounded rationality research programme}, initially proposed by Herbert Simon \cite{s1955} and systematically applied by Kahneman and Tversky \cite{kst1982,kt2000} to describe concrete judgements and decisions. Bounded rationality models give good predictions in a variety of circumstances. However, despite their simplicity and intuitive character, these models lack a unitary methodology, as well as deeper explanations, and thus provide a very fragmented picture of cognitive phenomena \cite{bbg2016}.

The {\it quantum cognition research programme} has recently attracted the interest of the scientific community due the superiority of quantum models over traditional and bounded rationality models to deal with the puzzles above. Quantum models were successfully applied to a variety of complex cognitive processes, including human language \cite{dc01,a2009,ags2013,dc02}, judgement \cite{bb2012,abgs2013,hk2013} and decision \cite{hk2013,ast2014,as2016} (see also \cite{s2017}). Despite these impressive results, however, quantum cognition still raises doubts regarding its explanatory power. Indeed, on the one side, quantum cognitive models introduce new parameters, which may fit experimental data, but do not necessarily explain them. On the other side, one is naturally led to wonder whether cognitive science really needs the entire mathematical formalism of quantum theory in Hilbert space or, on the contrary, non-Kolmogorovian non-Hilbertian models of probability are needed (see, e.g., \cite{holik2016,sergioli2017}).

In the present paper, we present binding motivations towards an independent, foundational, rather than purely phenomenological, justification of the quantum formalism in human judgement and decision-making under uncertainty. We start from the realistic and operational axiomatizations of quantum physics initiated by Jauch \cite{j1968} and Piron \cite{p1976} in Geneva and extended by Aerts (see, e.g., \cite{a1999,a2002}) in Brussels. Efforts have been made in the second part of the last century to derive the mathematical formalism of quantum theory in Hilbert space from more intuitive and empirically justified axioms, resting on basic notions directly connected with the operations that are performed in a laboratory. Particularly, in the Brussels approach, any physical entity is expressed in terms of the basic notions of state, context, property and mutual relationships between them ({\it SCoP} system). The approach is {\it realistic}, in the sense that the state, being the result of an effective preparation procedure, describes aspects of the reality of the entity. The approach is also {\it operational}, in the sense that all basic notions are expressed in terms of well defined empirical terms, like preparation and registration devices, statistics of outcomes, etc. If suitable ``purely operational'' axioms are imposed on a SCoP system, then the Hilbert space representation uniquely arises for the physical entity.

We believe that the above realistic-operational justification of the quantum Hilbert space formalism in physics also provides a strong motivation, if not a justification in itself, for the use of quantum Hilbert space structures in cognition. To this end we particularize in Sect. \ref{brussels} to a specific decision-making situation, the {\it Ellsberg paradox} situation, used as a case study here,  the realistic-operational foundation of cognitive entities we have recently elaborated \cite{asdbs2016}, in which a cognitive entity is abstractly described in terms of well defined empirical notions, i.e. state, context, property and outcome probability. Then, the stunning analogies in the realistic and operational descriptions of entities in physical and cognitive realms, suggest that the same Hilbert space leading axiomatics should be used for a cognitive, e.g., decision-making, entity (Sect. \ref{boundedrationality}).

The Ellsberg paradox is reviewed in Sect. \ref{ellsberg}, where we explain the difficulties of both expected utility and bounded rationality theories, to accommodate Ellsberg preferences and the results of more general Ellsberg-like decision situations.

We then elaborate in Sect. \ref{quantum} a mathematical representation in Hilbert space of the Ellsberg paradox situation and the ambiguity aversion pattern found in empirical literature. We had already presented quantum models of various Ellsberg thought experiments, including two-color and three-color urns \cite{ast2014,as2016,ahs2017,agms2018,s2019}. The novelty of the mathematical representation developed here consists in the fact that it follows directly from the canonical quantum representation of the realistic-operational terms of state, context, property and outcome probability in Hilbert space, which makes the use of quantum mathematics in this kind of situations more firmly founded and generalizable to other decision situations.

We finally offer some conclusive remarks and considerations in Sect. \ref{conclusions}, where we specify that the realistic-operational foundation of cognitive science can be in principle extended to several other judgement and decision-making situations, which constitutes a strong indication that ``possible failures of Hilbert space modelling" should be searched in other cognitive domains than individual judgements and decisions.

\section{Descriptive versus explanatory power of quantum structures\label{boundedrationality}}
Traditional theories of individual judgement and decision-making use, often implicitly, set-theoretic structures, that closely resemble the formal operations of classical Boolean logic and Kolmogorovian probability theory \cite{k1933}. This is specially evident in {\it rational decision theory}, according to which rational agents behave in such a way to maximize expected utility with respect to a Kolmogorovian probability measure and an underlying economic model \cite{vnm1944,s1954}.

These theories are normatively compelling, however, the judgement and decision puzzles in Sect. \ref{intro} make them descriptively problematical and suggest alternative more realistic approaches to human behaviour under uncertainty. A major research programme of this kind was initiated by Herbert Simon who put forward the {\it bounded rationality project} \cite{s1955}. Boundedly rational agents experience practical limitations in formulating and solving complex problems and in processing information. They tackle such limitations by taking mental short-cuts, making subjective evaluations and putting psychological aspects above rational reasoning. 

Within the bounded rationality project, one can cope with cognitive puzzles with judgement heuristics and reasoning biases, namely, the conjunction fallacy with the {\it representativeness heuristics} \cite{kst1982}, the Allais paradox with {\it prospect theory} \cite{kt2000}, the Ellsberg paradox with {\it cumulative prospect theory} \cite{kt2000}, the disjunction effect by {\it uncertainty aversion} \cite{kt2000}, etc. These approaches undoubtedly provide an intuitive account of how individuals actually behave in situations of uncertainty. However, the reader recognizes at once that a rather eclectic methodology or, better, a variety of methodologies,  are employed to accommodate the puzzles above and, while some authors support the hypothesis of an {\it adaptive toolbox} to deal with these problems \cite{gs2001}, many psychologists will find the bounded rationality research programme as unsatisfactory, while many  philosophers of science will try to derive these puzzling phenomena from a universal theory able to overcome the fragmentation of existing approaches.

The quantum cognition research programme reaches both effectiveness and unitarity. Since the nineties, quantum Hilbert space models have shown impressive superiority over traditional and bounded rationality approaches in dealing with the puzzles of human cognition and attributing them to genuine quantum effects, like contextuality, emergence, entanglement, interference and superposition. On the other side, quantum models introduce new parameters which can be possibly fitted by empirical data, without however necessarily explaining them. Hence, the quantum cognition research programme, though phenomenologically successful, does not seem to offer a deeper understanding and/or explanation of these puzzles. In addition, it is reasonable to wonder whether one really needs the entire Hilbert space formalism to represent cognitive phenomena, and should not better use more general non-Kolmogorovian representations outside physics (see, e.g., \cite{holik2016,sergioli2017}). In this respect, it should be noted that prospect theory already proposes non-Kolmogorovian probability models of probability of human decision \cite{kt2000}.

It is clear from the considerations above that one needs a {\it deeper justification} for the use of the Hilbert space formalism of quantum theory in cognition and decision and, more important, of its {\it necessity}. In this respect, a crucial result comes from increasing evidence that ``judgements and decisions create rather than record'' \cite{bb2012} -- see. e.g., the following quotations.

\begin{quote}
``There is a growing body of evidence that supports an alternative conception according to which preferences are often constructed – not merely revealed – in the elicitation process. These constructions are contingent on the framing of the problem, the method of elicitation, and the context of the choice.'' \cite{ts1993}
\end{quote}

\begin{quote} 
``\ldots the process of choice -- and in particular the act of choice -- can make substantial difference to what is chosen. \ldots, there is a particular necessity to take note of (i) chooser dependence, and (ii) menu dependence, of preference, even judged from a particular person's perspective.'' \cite{s1997}
\end{quote}

\begin{quote}
``\ldots valuations are initially malleable but become ‘imprinted’ after the agent is called upon to make an initial decision.'' \cite{apl2003}
\end{quote}

It is more and more acknowledged that, in any judgement or decision, a contextual interaction occurs between the situation that is the object of the evaluation and the individual who takes the decision (agent, decision-maker), which may affect the situation itself. At the end of this interaction, a result is actualized among a set of results that were only potential before the interaction \cite{ahs2017}. Hence, a judgement/decision process closely resembles a quantum measurement process, where a contextual interaction occurs between the  quantum particle that is measured and the measurement apparatus, which changes the state of the quantum particle determining what Heisenberg called ``transition from potential to actual''.

We believe that these analogies between micro-physics and cognition are a good starting point towards a foundational justification for the use of Hilbert space quantum formalism in cognition and decision.

In the sixties and seventies of the previous century, several authors wondered whether and how one can provide an independent justification for the Hilbert space formalism in quantum physics, deriving this formalism from physically justified axioms, resting on well defined empirical notions, directly connected with the operations that are usually performed in a laboratory. One of the well-known approaches to the foundations of quantum physics is the {\it Geneva-Brussels realistic-operational approach}, initiated by Jauch \cite{j1968} and Piron \cite{p1976} in Geneva, and successively extended by Aerts in Brussels (see, e.g., \cite{a1999,a2002}). This research consisted in abstractly describing any physical entity by relevant sets of states, contexts, properties and statistical connections between these notions (SCoP system). These theoretical notions are directly interpretable on physical operations on macroscopic apparatuses, such as preparation and registration devices, performed in spatio-temporal domains, such as physical laboratories. Measurements, state transformations, outcome probabilities and dynamics can then be expressed in terms of these more fundamental notions. If suitable axioms are imposed on the mathematical structures underlying a SCoP system, then the Hilbert space structure of quantum theory emerges as a unique mathematical representation, up to isomorphisms \cite{bc1981}. This justification provides the ``fundamental architecture of quantum theory in Hilbert space''.

We have recently proved that any cognitive entity $\Omega$, e.g., a concept, a conceptual combination, a proposition, or a more complex decision-making situation, can be abstractly described by a SCoP system $(\Sigma, {\cal \mathscr L}, {\mathscr C}, \mu, \nu)$ \cite{asdbs2016}, exactly like in physics. We review the essential elements of a SCoP system in cognition in the following.

(1) The complex of experimental procedures conceived by the experimenter, the experimental setting and the cognitive effect that one wants to analyse, define a {\it cognitive entity} $\Omega$, and are usually associated with a preparation procedure of a state of $\Omega$.

(2) $\Sigma$ is the set of all states of $\Omega$. A {\it state} $p$ of $\Omega$ is the consequence of a preparation procedure of $\Omega$ and has a cognitive, rather than physical, nature. The state of the cognitive entity is a {\it state of affairs}. It indeed expresses a ``reality of the cognitive entity'', in the sense that, once prepared in a given state, such condition is independent of any measurement procedure, and can be confronted with the different participants in an experiment, leading to outcome data and their statistics.

(3) ${\mathscr C}$ is the set of all contexts of $\Omega$. A {\it context} $e$ is an element that can provoke a change of state of the cognitive entity. A special context is the one introduced by a measurement. Indeed, when the cognitive experiment starts, an interaction occurs between the measured entity $\Omega$ under study and a participant in the experiment, in which the state $p$ of $\Omega$ generally changes, being transformed to another state $q$. This cognitive interaction is formalized by means of a context $e$.

(4) ${\cal \mathscr L}$ is the set of all properties of $\Omega$. A {\it property} $a$ of $\Omega$ is something $\Omega$ ``has'' independently of any context influencing the entity. An entity $\Omega$ is a given state $p$ has a set of properties that are {\it actual} in that state, the others being {\it potential}. A context $e$ may change the status actual/potential of a property, but cannot change the property itself.

(5) The change function $\mu: \Sigma \times {\mathscr C} \times \Sigma\longrightarrow [0,1]$ is such that, for every $p,q\in \Sigma$, $e\in {\mathscr C}$, $\mu(q,e,p)$ is the probability, as the large number limit of relative frequencies, that the context $e$ changes the initial state $p$ of $\Omega$ to the final state $q$.

Once recognizes at once in (1)--(5) the building blocks of the realistic-operational description of a physical entity, in the sense that in both physical and cognitive realms, a SCoP system incorporates all what is needed to study what an entity is, behaves and changes under a context. These impressive analogies indicate that the axioms generally used to justify the Hilbert space formalism of quantum physics are also appropriate to represent cognitive entities and processes. This provides an independent foundational clue and non-phenomenological motivation, if not a justification, that the mathematics of Hilbert space should be used to represent judgement and decision phenomena.

In Sect. \ref{brussels}, we will provide a realistic-operational description of a specific decision-making situation, the Ellsberg paradox, setting the grounds for a quantum mathematical representation of it in Sect. \ref{quantum}.  In the next section, we will instead summarize the serious difficulties of both traditional and bounded rationality approaches to handle such kind of decision-making situations.

\section{Rational decision theory and its puzzles\label{ellsberg}}
Traditional theories of rational decision-making rest on the tenet that, in situations of uncertainty, individual agents choose in such a way to maximize their expected utility, or degree of satisfaction.

In 1944, von Neumann and Morgenstern presented in a seminal work the first axiomatic formulation of expected utility theory. People continuously take decisions  among different options. These decisions are assumed to {\it reveal} underlying preferences. Then, von Neumann and Morgenstern proposed a set of ``reasonable'' axioms on human preferences such that, if the decisions are coherent, in the sense that they reveal axiom satisfying preferences, then the decisions are equivalent to the maximization of an expected utility functional with respect to a Kolmogorovian probability measure \cite{vnm1944}.

von Neumann and Morgenstern's formulation of expected utility theory has a major limitation, in that it only deals with the uncertainty that can be formalized by known probabilities (also referred to as {\it objective uncertainty}, or {\it risk}). On the other hand, situations frequently occur in which uncertainty cannot be formalized by known probabilities (also referred to as {\it subjective uncertainty}, or {\it ambiguity}) \cite{k1921}. The Bayesian approach to probability minimizes the distinction between objective and subjective uncertainty introducing the notion of {\it subjective probability}. Even when probabilities are not known, people may still construct their own {\it beliefs}, or {\it priors} (which may differ from one individual to another), and they maximize expected utility with respect to these priors. Indeed, Leonard Savage presented in 1954 an axiomatic formulation of expected utility theory which extends the one of von Neumann and Morgenstern to subjective uncertainty \cite{s1954}.

We summarize in the following the essential definitions and results of Savage's expected utility theory, together with its major pitfalls. We refer to \cite{gm2013,ms2014} for detailed reviews of these results.

Savage introduced a set of basic notions, including states of nature, consequences, preferences, and looked for justified axioms on preferences  able to provide a representation theorem in which ordering of preferences is characterized by maximization of expected utility. This procedure formally resembles the procedures used in the axiomatizations of quantum physics in Sect. \ref{brussels}.

Let ${\mathscr S}$ be the set of all (physical) {\it states of nature}, which we assume to be discrete and finite here, for the sake of simplicity. Let ${\mathscr P}({\mathscr S})$ be the power set of ${\mathscr S}$ and ${\mathscr A}\subseteq {\mathscr P}({\mathscr S})$ be a (Boolean) $\sigma$-algebra. An element $E\in {\mathscr A}$ denotes an {\it event}. A {\it Kolmogorovian probability measure} over $\mathscr A$ is a function $p:{\mathscr A}\subseteq {\mathscr P}({\mathscr S})\longrightarrow [0,1]$ satisfying the axioms of Kolmogorov \cite{k1933}.

Then, let ${\mathscr X}$ be the set of all {\it consequences}, whose elements we assume to denote {\it monetary payoffs}, hence real numbers, here, for the sake of simplicity. In Savage's formulation, a function $f: {\mathscr S} \longrightarrow {\mathscr X}$ denotes an {\it act}. Let $\mathscr{F}$ be the set of all acts. Let us endow $\mathscr{F}$ with a {\it weak preference relation} $\succsim$, that is, a reflexive, symmetric and transitive relation over the Cartesian product $\mathscr{F} \times \mathscr{F}$. In $\succsim$, the relations $\succ$ and $\sim$ denote {\it strong preference} and {\it indifference}, respectively, that is, we write $f \succ g$ whenever an individual strictly prefers act $f$ to act $g$ and $f \sim g$ whenever the individual is indifferent between $f$ and $g$.

Next, let $u: {\mathscr X} \longrightarrow \Re$ be a {\it utility function} over ${\mathscr X}$. This function typically expresses the decision-maker's taste, hence it is assumed to be strictly increasing and continuous, with additional technical constraints related to the specification of the decision-maker's attitude towards risk.

The mathematical definitions above can be simplified by introducing a set $\{E_1, \ldots, E_n \}$ of mutually exclusive and exhaustive elementary events, where $E_i=\{ s_i \in {\mathscr S}\}$, $i\in \{1,\ldots, n\}$, which thus form a partition of ${\mathscr S}$. For every $i \in \{ 1, \ldots, n\}$, let $x_i$ be the utility associated by the act $f$ to the event $E_i$. Then, $f$ can be equivalently expressed by the 2n-tuple $f=(E_1,x_1;\ldots;E_n,x_n)$, meaning that the individual will get the outcome $x_1$ if the event $E_1$ occurs (i.e. the state of nature $s_1$ realizes), \ldots, the outcome $x_n$ if the event $E_n$ occurs (i.e. the state of nature $s_n$ realizes).

Finally, we denote by $W(f)=\sum_{i=1}^{n}p(E_i)u(x_i)$ the {\it expected utility} associated with the act $f$ with respect to the Kolmogorovian probability measure $p$. 

In his representation theorem, Savage proved that, if the algebraic structure $({\mathscr F}, \succsim)$ satisfies a number of ``reasonable'' axioms\footnote{One of the axioms is the famous {\it sure thing principle}, which is violated in the Ellsberg paradox. The other axioms are: ordinal event independence, comparative probability, non-degeneracy, small event continuity and dominance, and have a technical nature \cite{s1954}. However, these axioms are not relevant to the present purposes, hence we will not dwell on them, for the sake of brevity.} then, for every $f,g\in \mathscr F$, a unique Kolmogorovian probability measure $p$ and a unique (up to positive affine transformations) utility function $u$ exist such that $f$ is preferred to $g$, i.e. $f \succsim g$, if and only if the expected utility of $f$ is greater than the expected utility of $g$, i.e. $W(f) \ge W(g)$. For every $i\in \{ 1, \ldots, n\}$, the utility value $u(x_i)$ depends on the decision-maker's risk preferences, while $p(E_i)$ is interpreted as the subjective probability, expressing the individual's belief that the event $E_i$ occurs \cite{s1954}.

Savage's result is both compelling at a {\it normative level} and testable at a {\it descriptive level}. Indeed:

(i) if the axioms are intuitively reasonable and decision-makers agree with them, then they {\it must} all behave as if they were maximizing an expected utility with respect to a single subjective probability distribution satisfying Kolmogorov's axioms;

(ii) the axioms suggest to design decision-making experiments to test the validity of expected utility theory, hence of the axioms themselves, in real life situations.

Because of (i), Savage's expected utility formulation is generally accepted to prescribe ``how rational agents should choose''. However, one the one side, the theory offers very little about where beliefs come from and how they should be calculated and, on the other side, regarding (ii), decision-making experiments have systematically found deviations from that rational behaviour in concrete situations. 

In particular, Daniel Ellsberg proved in 1961 in a number of thought experiments that decision-makers generally prefer acts with known (or objective) probabilities over acts with unknown (or subjective) probabilities \cite{e1961}. We analyse here the famous {\it Ellsberg three-color example} as a paradigmatic example to show that (i) traditional decision theories do not work, (ii) bounded rationality approaches are not sufficiently explanatory, (iii) quantum structures are  needed.

Consider one urn with 30 red balls and 60 balls that are either yellow or black, the latter in unknown proportion. One ball will be drawn at random from the urn. Then, free of charge, a person is asked to bet on pairs of the acts $f_1$, $f_2$, $f_3$ and $f_4$ in Table 1. 
\noindent 
\begin{table} \label{table01}
\begin{center}
\begin{tabular}{|p{1.5cm}|p{1.5cm}|p{1.5cm}|p{1.5cm}|}
\hline
\multicolumn{1}{|c|}{} & \multicolumn{1}{c|}{1/3} & \multicolumn{2}{c|}{2/3} \\
\hline
Act & Red & Yellow & Black \\ 
\hline
\hline
$f_1$ & \$100 & \$0 & \$0 \\ 
\hline
$f_2$ & \$0 & \$0 & \$100 \\ 
\hline
$f_3$ & \$100 & \$100 & \$0 \\ 
\hline
$f_4$ & \$0 & \$100 & \$100 \\ 
\hline
\end{tabular}
\end{center}
{\bf Table 1.} The payoff matrix for the Ellsberg three-color example.
\end{table}
\noindent 
Ellsberg suggested that, when asked to rank these acts, most individuals will prefer  $f_1$ over $f_2$ and $f_4$ over $f_3$. Indeed, $f_1$ and $f_4$ are {\it unambiguous acts}, in the sense that they are associated with events over known probabilities -- the events ``a red ball is drawn'' and ``a yellow or black ball is drawn'' are associated with objective probabilities 1/3 and 2/3, respectively. On the contrary, $f_2$ and $f_3$ are {\it ambiguous acts}, in the sense that they are associated with events over unknown probabilities -- the events ``a yellow ball is drawn'' and ``a black ball is drawn'' are both associated with a probability ranging from 0 to 2/3. This attitude of decision-makers to prefer ``probabilized over non-probabilized uncertainty'' has been known as {\it ambiguity aversion} since Ellsberg studies \cite{e1961}.

Several experiments on Ellsberg urns decisions, but also on financial, insurance and medical decisions, have confirmed the {\it Ellsberg preferences} $f_1 \succ f_2$ and $f_4 \succ f_3$, thus indicating that ambiguity aversion is a good candidate to explain concrete decisions in this case, and only Slovic and Tversky found {\it ambiguity seeking patterns} (see, e.g., \cite{ms2014} for a review of experimental studies).

In \cite{agms2018}, we tested various human decision puzzles, including the Ellsberg three-color example. We asked 200 people, chosen among colleagues and friends, to fill a questionnaire in which they had to choose between various options. People had on average a basic knowledge of probability theory, but no specific training in decision theory. Participants were provided with a questionnaire similar to the one in Figure 1, in which they had to choose between acts $f_1$ and $f_2$ and, then, between acts $f_3$ and $f_4$ in Table 1.
\begin{figure}
\begin{center}
\includegraphics[scale=0.6]{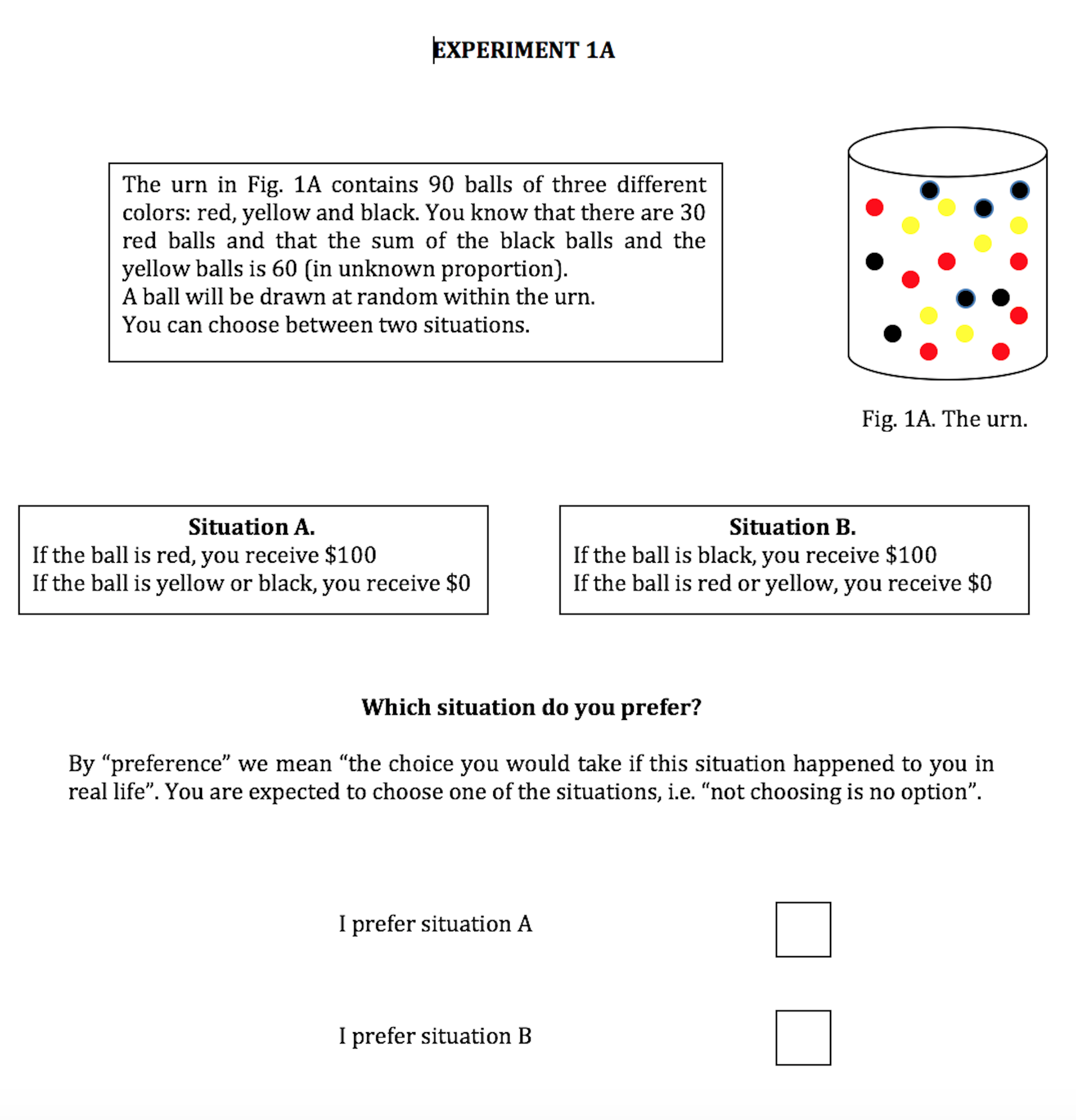}
\end{center}
{\bf Figure 1.} A sample of the questionnaire related to the decision-making experiment on the Ellsberg three-color example: choice between acts $f_1$ and $f_2$ in Table 1.
\end{figure}
Overall, 125 participants preferred acts $f_1$ and $f_4$, 38 preferred acts $f_1$ and $f_3$, 6 preferred acts $f_2$ and $f_3$, and 31 preferred acts $f_2$ and $f_4$. This means that 163 participants over 200 preferred act $f_1$ over act $f_2$, which entails a {\it preference weight} of 0.815. Also, 156 participants over 200 preferred act $f_4$ over act $f_3$, which entails a preference weight of 0.780. The inversion rate is 0.655, a pattern that agrees with the Ellsberg preferences found in the literature and significantly indicates the presence of ambiguity aversion.

Preferences of decision-makers who are sensitive to ambiguity, that is, are ambiguity averse or ambiguity seeking, cannot be explained within Savage's expected utility theory, because they violate the sure thing principle, according to which, preferences should be independent of the common outcome. In the specific case of the three-color example, preferences should not depend on whether the common event ``a yellow ball is drawn'' pays off \$0 or \$100. More technically, Savage's expected utility theory predicts {\it consistency of preferences}, namely, $f_1$ is preferred to $f_2$ if and only if $f_3$ is preferred to $f_4$. A simple calculation shows that this is impossible within a traditional expected utility framework. Indeed, if we denote by $\tilde{p}_{R}$, $\tilde{p}_{Y}$ and $\tilde{p}_{B}$ the probability that a red ball, a yellow ball, a black ball, respectively, is drawn (with $\tilde{p}_{R}=1/3=1-(\tilde{p}_{Y}+\tilde{p}_{B})$), then the expected utilities $W(f_{i})$, $i=1,2,3,4$, are such that $W(f_1)>W(f_2)$ if and only if $(\tilde{p}_{R}-\tilde{p}_{B})(u(100)-u(0))>0$ if and only if $W(f_3)>W(f_4)$. We can equivalently say that no assignment of Kolmogorovian probabilities $\tilde{p}_{R}$, $\tilde{p}_{Y}$ and $\tilde{p}_{B}$ reproduces a preference with $W(f_1)>W(f_2)$  and $W(f_4)>W(f_3)$, whence the {\it Ellsberg paradox}. 

Several extensions of Savage's expected utility theory have been put forward in order to accommodate the Ellsberg paradox (see, e.g., the reviews in \cite{gm2013} and \cite{ms2014}). One of the major proposals is Tversky and Kahneman's cumulative prospect theory, mentioned in Sect. \ref{boundedrationality} and elaborated within a bounded rationality research programme \cite{kt2000}. In particular, to reproduce Ellsberg preferences, Tversky and Kahneman replaced (i) the utility function $u$ by a {\it scaling function} $u'$ reflecting the subjective value of the outcome utility, and (ii) the subjective probability measure $p$ by a non-additive measure $p'$ satisfying the mathematical properties of a {\it capacity}. As we have mentioned in Sect. \ref{boundedrationality}, such bounded rationality models, though descriptively interesting and easily interpretable intuitively, provide a too fragmented view of decision theory, hence they are not able to provide a unitary and adequate explanatory framework to understand the deep aspects of decision processes. In addition, cumulative prospect theory, as well as other major non-expected utility models, fails to reproduce the empirical results of a recently elaborated variant of the Ellsberg paradox, the {\it Machina paradox} \cite{agms2018,m2009,lhp2010}.

An innovative aspect of descriptive, like bounded rationality, approaches, is the representation of subjective probabilities by more general, possibly non-Kolmogorovian, mathematical structures. This is crucial towards a more satisfactory framework for human decision-making that goes beyond Savage's expected utility, as we will see in Sect. \ref{quantum}.

In the next section, we intend to elaborate a realistic-operational description of a decision-making situation, using the Ellsberg  three-color example as a case study. We will demonstrate that, once the Ellsberg paradox situation is formulated in terms of states, contexts, properties and transition probabilities, then the application of the mathematical formalism of quantum theory directly follows from the canonical representation of these realistic-operational terms in Hilbert space.

\section{A realistic-operational description of a decision-making situation\label{brussels}}
In this section, we specify the realistic-operational description of cognitive entities in Sect. \ref{boundedrationality} to the decision-making situation presented in the Ellsberg three-color example \cite{asdbs2016}. In it, we explicitly distinguish physical from cognitive, in this case, decision-making, entities. Analogously, we distinguish physical from cognitive states of nature, though one can intuitively see that some cognitive states are mapped into the corresponding physical states.

In the Ellsberg three-color example, the cognitive, i.e. decision-making, entity $\Omega_{DM}$ is the urn with 30 red balls and 60 yellow or black balls in unknown proportion. This is what the individual reads in a questionnaire, interacts with and takes a decision on.

The cognitive entity $\Omega_{DM}$ is associated with a defined set $\Sigma_{DM}$ of states.\footnote{Some authors identify the notion of ``state'' with the notion of ``belief state'' of the individual participating in the cognitive experiment, e.g., taking the decision (see, e.g., \cite{bb2012,bbg2016,hk2013}). We instead neatly distinguish states from measurements here. A state is defined by a preparation procedure of the cognitive entity under investigation. The participant in the experiment acts as a (measurement) context that interacts with the cognitive entity and changes its state.} A state $p \in \Sigma_{DM}$ has a cognitive nature and incorporates aspects of ambiguity. A context $e$ does not pertain to  $\Omega_{DM}$ but can interact with it. Let ${\mathscr C}_{DM}$ be the set of all contexts of $\Omega_{DM}$. The interaction of $\Omega_{DM}$ with a context $e \in {\mathscr C}_{DM}$ may determine a change of the state of $\Omega_{DM}$ from $p$ to a different state $q$. The probability of such a state transition will be denoted by $\mu(q,e,p)$, where $\mu:\Sigma_{DM} \times {\mathscr C}_{DM} \times \Sigma_{DM} \longrightarrow [0,1]$. We might complete the realistic-operational description of $\Omega_{DM}$ defining a set ${\mathscr L}_{DM}$ of properties and an actuality relation connecting properties and states. However, they are not needed in the Ellsberg three-color scenario, hence we omit specifying these notions here, for the sake of brevity, though they may be relevant in more general decision situations.

Let us now introduce a color context $e_{C} \in {\mathscr C}_{DM}$, which is the context associated with a drawing of a ball from the urn. As a result of the drawing, we have three possible outcomes, $R$, $Y$ and $B$, corresponding to the colors of the balls, red, yellow and black, respectively. The outcomes $R$, $Y$ and $B$ are the eigenvalues of $e_{C} $ and are respectively associated with the final states, or eigenstates, $p_{R}$, $p_{Y}$ and $p_{B}$ of the cognitive entity $\Omega_{DM}$. These eigenstates are such that $\mu(p_{i},e_{C},p_{i})=1$, $i \in \{R,Y,B\}$.

The color context $e_{C}$ introduces three mutually exclusive and exhaustive elementary events $E_{i}=(e_{C}, \{ i\})$, $i \in \{R,Y,B\}$, which are such that the subjective probability that the event $E_i$ occurs when the cognitive entity $\Omega_{DM}$ is in the state $p$ is given by the transition probability $\mu(p_{i}, e_{C}, p)\in [0,1]$, $i \in \{R,Y,B\}$.

Then, in analogy with Savage's expected utility theory, we can introduce monetary payoffs $x \in {\mathscr X} \subseteq \Re$, utility functions $u:{\mathscr X}\longrightarrow \Re$ and acts taking the form $f=(E_R,x_R;E_Y,x_Y,E_B,x_B)$, mapping the event $E_i$ into the payoff $x_i$, $i \in \{R,Y,B\}$. In particular, the acts $f_1,f_2,f_3$ and $f_4$ in Table 1, Sect. \ref{ellsberg} are defined in the way above.

Let us now come to the operational-realistic description of a decision-making process in the Ellsberg three-color situation. Suppose that, in the absence of any context, the entity $\Omega_{DM}$ is in the initial state $p_0$. This state corresponds to a preparation of the cognitive entity $\Omega_{DM}$ and can be set by the information on the corresponding physical entity. For example, it is reasonable to assume that $p_0$ is such that, for every $i \in \{ R,Y,B\}$, $\mu(p_i, e_{C},p_0)=1/3$, because of the indifference principle. Whenever an individual is asked to rank $f_1$ and $f_2$,  the individual's attitude towards ambiguity, e.g., ambiguity aversion, can be described as a new context $e_1 \in {\mathscr C}_{DM}$ acting on $\Omega_{DM}$ in the initial state $p_0$ and changing $p_0$ into a new state $p_1$, characterized by a new probability distribution $\mu(p_{i},e_{C},p_1)$, $i \in \{R,Y,B\}$. Similarly, whenever the individual is asked to rank $f_3$ and $f_4$,  the individual's attitude towards ambiguity, e.g., ambiguity aversion, can be described as a new context $e_2 \in {\mathscr C}_{DM}$ acting on $\Omega_{DM}$ in the initial state $p_0$ and changing $p_0$ into a new state $p_2$, characterized by a new probability distribution $\mu(p_{i},e_{C},p_2)$, $i \in \{R,Y,B\}$. The cognitive states $p_1$ and $p_2$, and their ensuing probability distributions, are responsible of the {\it inversion of preferences} which occur in the Ellsberg paradox situation.

Finally, a decision process between acts $f_1$ and $f_2$ can be operationally described as a measurement context $e_{12}\in {\mathscr C}_{DM}$ acting on the entity $\Omega_{DM}$ in the ambiguity averse state $p_1$, with possible outcomes ``yes'' and ``no''. Similarly, a decision process between acts $f_3$ and $f_4$ can be described as a measurement context $e_{34}\in {\mathscr C}_{DM}$ acting on the entity $\Omega_{DM}$ in the ambiguity averse state $p_2$, with possible outcomes ``yes'' and ``no''. These contexts give rise of the statistics of outcomes in a decision-making test on the Ellsberg three-color urn.

Now, the considerations in Sect. \ref{boundedrationality} naturally indicate to represent states, contexts, properties and outcome probabilities of $\Omega_{DM}$ by using the canonical quantum representation of states, contexts, properties and outcome probabilities in Hilbert space. In particular, subjective probabilities will be represented using the Born rule of quantum probability. This is what we intend to show in the next section where the realistic-operational terms defined here will be canonically represented using quantum mathematical terms.

\section{A novel quantum representation of the Ellsberg paradox\label{quantum}}
In this section we elaborate a new quantum representation of the three-color example straightly following the canonical  Hilbert space representation of the realistic and operational notions in Sect. \ref{ellsberg}. This representation generalizes and strengthens those in \cite{ast2014,as2016,ahs2017,agms2018}.

\subsection{Quantum representation of basic notions}
The cognitive entity $\Omega_{DM}$ is associated with a Hilbert space $\mathscr H$. Since, the three-color example involves three mutually exclusive and exhaustive elementary events, $\mathscr H$ can be chosen to be isomorphic to the complex Hilbert space $\mathbb{C}^{3}$ of ordered triples of complex numbers. Let $\{(1,0,0),(0,1,0),(0,0,1) \}$ be the canonical orthonormal basis of $\mathbb{C}^{3}$.

A state $p_{v}$ of the entity $\Omega_{DM}$ is represented by the unit vector $|v \rangle \in {\mathscr H}$, $|| |v\rangle||=\sqrt{\langle v|v\rangle}=1$.

The elementary event $E_{i}$ is represented by the one-dimensional orthogonal projection operator $P_{i}=|i\rangle\langle i|$, $i \in \{R,Y,B\}$, where we choose $|R\rangle=(1,0,0)$, $|Y\rangle=(0,1,0)$ and $|B\rangle=(0,0,1)$. The color context $e_{C}$ is then represented by the spectral family $\{P_{R}, P_{Y}, P_{B} \}$.

In the canonical basis of $\mathbb{C}^{3}$, the unit vector $|v \rangle$ can be written as
\begin{equation} \label{vector_p_v}
|v \rangle=\rho_{R}e^{i \theta_R}|R\rangle+\rho_{Y}e^{i \theta_Y}|Y\rangle+\rho_{B}e^{i \theta_B}|B\rangle=(\rho_{R}e^{i \theta_R},\rho_{Y}e^{i \theta_Y},\rho_{B}e^{i \theta_B})
\end{equation}

We use the Born rule to represent subjective probabilities. Then, the subjective probability that the elementary event $E_i$, $i\in \{ R,Y,B\}$, occurs when the entity $\Omega_{DM}$ is in the state $p_{v}$ is given by
\begin{equation}
\mu_{v}(E_i)=\langle v |P_{i}|v\rangle=|\langle i|v\rangle|^{2}=\rho_{i}^{2}
\end{equation}
In addition, the subjective probability that the event $E$, represented by the orthogonal projection operator $P_E$, occurs when the cognitive entity $\Omega_{DM}$ is in the state $p_{v}$ is $\mu_{v}(E)=\langle v |P_E|v\rangle=||P_E|v\rangle||^{2}$. Finally, for every state $p_{v}$ represented by the unit vector $|v\rangle$, the subjective probability measure
\begin{equation}
\mu_{v}:{\mathscr L}(\mathbb{C}^{3})\longrightarrow [0,1]
\end{equation}
is a quantum probability measure over the lattice ${\mathscr L}(\mathbb{C}^{3})$ of all orthogonal projection operators on the Hilbert space $\mathbb{C}^{3}$.

Compatibility with the standard three-color situation entails $\rho_{R}^{2}=\frac{1}{3}$, hence it follows from Eq. (\ref{vector_p_v}) that
\begin{equation}
|v \rangle=(\frac{1}{\sqrt{3}}e^{i \theta_R},\rho_{Y}e^{i \theta_Y},\rho_{B}e^{i \theta_B})=(\frac{1}{\sqrt{3}}e^{i \theta_R},\rho_{Y}e^{i \theta_Y},\sqrt{\frac{2}{3}-\rho_{y}^{2}}e^{i \theta_B})
\end{equation}
where the last equality is obtained from $\sqrt{\langle v|v\rangle}=1$.

Special states are the state with no black balls represented by
\begin{equation}
|v_{RY} \rangle=(\frac{1}{\sqrt{3}}e^{i \theta_R},\sqrt{\frac{2}{3}}e^{i \theta_Y},0)
\end{equation}
and the state with no yellow balls represented by
\begin{equation}
|v_{RB} \rangle=(\frac{1}{\sqrt{3}}e^{i \theta_R},0,\sqrt{\frac{2}{3}}e^{i \theta_B})
\end{equation}
The acts $f_1$, $f_2$, $f_3$ and $f_4$ are represented by the self-adjoint operators
\begin{eqnarray}
\hat{F_1}&=&u(100)P_{R}+u(0)P_Y+u(0)P_B \\
\hat{F_2}&=&u(0)P_{R}+u(0)P_Y+u(100)P_B \\
\hat{F_3}&=&u(100)P_{R}+u(100)P_Y+u(0)P_B \\
\hat{F_4}&=&u(0)P_{R}+u(100)P_Y+u(100)P_B
\end{eqnarray}
respectively. The utility function $u(\cdot)$ is not given by the theory but it is revealed in a decision test by concrete choices, in analogy with standard procedures in the literature.

The expected utility $W_{v}(f_i)$, $i\in \{1,2,3,4\}$, in a generic state $p_{v}$ of the entity $\Omega_{DM}$ is
\begin{eqnarray}
W_{v}(f_1)&=&\langle v|\hat{F_1}|v \rangle=\frac{1}{3}u(100)+\frac{2}{3}u(0) \\
W_{v}(f_2)&=&\langle v|\hat{F_2}|v \rangle=(\frac{1}{3}+\rho_{Y}^{2})u(0)+(\frac{2}{3}-\rho_{Y}^{2})u(100) \\
W_{v}(f_3)&=&\langle v|\hat{F_3}|v \rangle=(\frac{1}{3}+\rho_{Y}^{2})u(100)+(\frac{2}{3}-\rho_{Y}^{2})u(0) \\
W_{v}(f_4)&=&\langle v|\hat{F_4}|v \rangle=\frac{1}{3}u(0)+\frac{2}{3}u(100)
\end{eqnarray}
As we can see, the expected utilities $W_{v}(f_1)$ and $W_{v}(f_4)$ do not depend on the cognitive state $p_{v}$ of the entity $\Omega_{DM}$, in agreement with the fact that $f_1$ and $f_4$ are unambiguous acts. On the contrary, the expected utilities $W_{v}(f_2)$ and $W_{v}(f_3)$ do depend on the cognitive state $p_{v}$, in agreement with the fact that $f_2$ and $f_3$ are ambiguous acts. This also agrees with our assumption that cognitive states provide information on ambiguity.

\subsection{Reproducing Ellsberg preferences with ambiguity averse states}
Let us now suppose that, in the absence of any context, the cognitive entity $\Omega_{DM}$ is in the initial state $p_{0}$. The principle of indifference (see Sect. \ref{brussels}) then suggests that $p_{0}$ is the state represented by the unit vector
\begin{equation}
|v_{0} \rangle=\frac{1}{\sqrt{3}}(1,1,1)
\end{equation}
leading to uniform probabilities of drawing a red, yellow and black ball. The ambiguity attitude contexts $e_{1}$ and $e_{2}$ will determine a change of state of the entity $\Omega_{DM}$, depending on individual preferences towards ambiguity. For example, two ambiguity seeking states $p_{w_{1}}$ and $p_{w_{2}}$ will be such that the following inequalities hold
\begin{eqnarray}
W_{w_{1}}(f_2)&>&\frac{1}{3}u(100)+\frac{2}{3}u(0) \\
W_{w_{2}}(f_3)&>&\frac{1}{3}u(0)+\frac{2}{3}u(100)
\end{eqnarray}

We will instead explicitly determine two ambiguity averse states $p_{w_{1}}$ and $p_{w_{2}}$ which reproduce Ellsberg preferences, that is,
\begin{eqnarray}
W_{w_{1}}(f_1)&>&W_{w_{1}}(f_2) \\
W_{w_{2}}(f_4)&>&W_{w_{2}}(f_3)
\end{eqnarray}
Two general cognitive states $p_{w_{1}}$ and $p_{w_{2}}$  states are represented by the unit vectors
\begin{eqnarray}
|w_{1} \rangle&=&(\frac{1}{\sqrt{3}}e^{i \theta_R},\rho_{Y}e^{i \theta_Y},\sqrt{\frac{2}{3}-\rho_{y}^{2}}e^{i \theta_B}) \\
|w_{2} \rangle&=&(\frac{1}{\sqrt{3}}e^{i \phi_R},\tau_{Y}e^{i \phi_Y},\sqrt{\frac{2}{3}-\tau_{y}^{2}}e^{i \phi_B})
\end{eqnarray}
respectively. For the sake of simplicity, let us look for states with simple phases, namely, $\theta_{R}=\phi_{R}=0$ and $\theta_{Y},\theta_{B},\phi_{Y},\phi_{B}=0,\pi$. In particular, one can show that, for every $\alpha>\frac{1}{\sqrt{3}}$, the unit vectors
\begin{eqnarray}
|w_{1} \rangle&=&(\frac{1}{\sqrt{3}},\alpha,-\sqrt{\frac{2}{3}-\alpha^{2}}) \\
|w_{2} \rangle&=&(\frac{1}{\sqrt{3}},-\sqrt{\frac{2}{3}-\alpha^{2}},\alpha)
\end{eqnarray}
reproduce Ellsberg preferences. However, the ambiguity averse states $p_{w_{1}}$ and $p_{w_{2}}$ are not generally orthogonal, unless $\alpha=\pm 0.7887$. Let us choose the positive sign, so that the orthonormal vectors
\begin{eqnarray}
|w_{1} \rangle&=&(0.5774,0.7887,-0.2113) \label{w1} \\
|w_{2} \rangle&=&(0.5774,-0.2113,0.7887) \label{w2}
\end{eqnarray}
reproduce Ellsberg preferences in the three-color example within a quantum mathematical representation.

\subsection{Modelling empirical data in Hilbert space}
The final step of the quantum representation of the three-color example consists in modelling the experimental data in Sect. \ref{ellsberg}. To this aim, we describe the decision between acts $f_1$ and $f_2$ by a measurement context $e_{12}$ and represent the latter by the spectral family $\{ M, \mathbbmss{1}-M \}$, where the orthogonal projection operator $M$ projects onto the one-dimensional subspace generated by the unit vector $|m\rangle=(\frac{1}{\sqrt{3}}, \rho_{Y}e^{i \theta_Y}, \rho_{B}e^{i \theta_B})$, where $\rho_Y^2+\rho_B^2=\frac{2}{3}$. The one-dimensional projection operator $M$ is then given by
\begin{equation} \label{projM}
M=|m\rangle \langle m|=\left( 
\begin{array}{ccc}
 \frac{1}{3} & \frac{1}{\sqrt{3}}\rho_Y e^{-i\theta_Y} & \frac{1}{\sqrt{3}}\rho_B e^{-i\theta_B}\\
 \frac{1}{\sqrt{3}}\rho_Y e^{i\theta_Y} & \rho_Y^2 & \rho_Y \rho_B e^{i (\theta_Y-\theta_B)}  \\
\frac{1}{\sqrt{3}}\rho_B e^{i\theta_B} & \rho_Y \rho_B e^{-i (\theta_Y-\theta_B)} & \rho_B^2
\end{array} \right)
\end{equation}
Analogously, we describe the decision between acts $f_3$ and $f_4$ by a measurement context $e_{34}$ and represent the latter by the spectral family $\{ N, \mathbbmss{1}-N \}$, where the orthogonal projection operator $N$ projects onto the one-dimensional subspace generated by the unit vector $|n\rangle=(\frac{1}{\sqrt{3}}, \tau_{Y}e^{i \phi_Y}, \tau_{B}e^{i \phi_B})$, where $\tau_Y^2+\tau_B^2=\frac{2}{3}$. The one-dimensional projection operator $N$ is then given by
\begin{equation} \label{projN}
N=|n\rangle \langle n|=\left( 
\begin{array}{ccc}
 \frac{1}{3} & \frac{1}{\sqrt{3}}\tau_Y e^{-i\phi_Y} & \frac{1}{\sqrt{3}}\tau_B e^{-i\phi_B}\\
 \frac{1}{\sqrt{3}}\tau_Y e^{i\phi_Y} & \tau_Y^2 & \tau_Y \tau_B e^{i (\phi_Y-\phi_B)}  \\
\frac{1}{\sqrt{3}}\tau_B e^{i\phi_B} & \tau_Y \tau_B e^{-i (\phi_Y-\phi_B)} & \tau_B^2
\end{array} \right)
\end{equation}
We refer to the unit vectors $|v_0\rangle=\frac{1}{\sqrt{3}}(1,1,1)$ and $|w_1\rangle$ and $|w_2\rangle$ in Eqs. (\ref{w1})
and (\ref{w2}). It follows that the conditions
\begin{eqnarray}
\langle m|m\rangle&=&1 \label{norm_m} \\
\langle n|n\rangle&=&1 \label{norm_n} \\
\langle w_1|M|w_1\rangle&=&0.815 \label{exp12} \\
\langle w_2|N|w_2\rangle&=&0.780 \label{exp34} \\
\langle v_0|M|v_0\rangle&=&0.500 \label{sym12} \\
\langle v_0|N|v_0\rangle&=&0.500 \label{sym34}
\end{eqnarray}
must be satisfied by the real parameters $\rho_Y$, $\rho_B$, $\theta_Y$, $\theta_B$, $\tau_Y$, $\tau_B$, $\phi_Y$ and $\phi_B$. Equations (\ref{norm_m}) and (\ref{norm_n}) are determined by normalization conditions, while Eqs. (\ref{exp12}) and (\ref{exp34}) are determined by empirical data. Finally, Eqs. (\ref{sym12}) and (\ref{sym34}) are determined by the fact that decision-makers who are not sensitive to ambiguity should overall be indifferent between $f_1$ and $f_2$, as well as between $f_3$ and $f_4$. Hence, on average, half respondents are expected to prefer $f_1$ ($f_3$) and the other half $f_2$ ($f_4$). To simplify the analysis, let us set $\theta_Y=\phi_B=0$. Hence, we are left with a system of 6 equations in 6 unknown variables whose solution is
\begin{eqnarray}
\left\{ \begin{array}{lll}    
\rho_Y&=&0.6853 \\
\rho_B&=&0.4438 \\
\theta_B&=&105.07^{\circ} \\
\tau_Y&=&0.4785 \\
\tau_B&=&0.6616 \\
\phi_Y&=&102.87^{\circ}
 \end{array} \right.
\end{eqnarray}
Equivalently, we get
\begin{eqnarray}
|m\rangle&=&(0.5773,0.6853,0.4438 e^{i 105.07^{\circ}}) \\
|n\rangle&=&(0.5773,0.4785 e^{i 102.87^{\circ}}, 0.6616)
\end{eqnarray}
Thus, the orthogonal projection operators in Eqs. (\ref{projM}) and (\ref{projN}) reproducing the experimental data in Sect. \ref{ellsberg} are
\begin{eqnarray}
M&=&\left(
\begin{array}{ccc}
0.333 & 0.396 & 0.256 e^{-i 105.07^{\circ}}\\
0.396 & 0.470 & 0.304 e^{-i 105.07^{\circ}} \\
0.256 e^{i 105.07^{\circ}} & 0.304 e^{i 105.07^{\circ}} & 0.197
\end{array} 
\right) \\
N&=&\left(
\begin{array}{ccc}
0.333 & 0.276 e^{-i 102.87^{\circ}} & 0.382\\
0.276 e^{i 102.87^{\circ}}  & 0.229 & 0.317 e^{i 102.87^{\circ}} \\
0.382 & 0.317 e^{-i 102.87^{\circ}} & 0.438
\end{array} 
\right)
\end{eqnarray}
The construction of a quantum model for the data on the Ellsberg three-color experiment in Sect. \ref{ellsberg} is thus completed. As we can see, the quantum model naturally arises from the canonical Hilbert space representation of the realistic-operational termsin Sect. \ref{brussels}.

\section{Conclusions\label{conclusions}}
Despite its phenomenological success to deal with classically problematical cognitive puzzles, the quantum cognition research programme still poses challenging questions regarding its explanatory power and necessity.

In this paper, we specialized to the Ellsberg paradox decision situation a realistic-operational foundation which we have recently extended  from physics to cognition. Then, we applied to the Ellsberg three-color example the canonical quantum representation of realistic-operational terms in Hilbert space.

This result on the Ellsberg paradox situation is paradigmatic, in the sense that one can follow the same strategy to generally claim that the mathematical representation of human judgements and decision-making in Hilbert space has now an independent motivation of a foundational, rather than phenomenological, nature.

To conclude, we agree that quantum theory in Hilbert space is not the ultimate theory in cognition -- recent results on sequential measurements and order effects seem to confirm this conclusion(see, e.g., \cite{asdbs2016}). However, we also believe that there are strong theoretical motivations, in addition to its empirical success and unitary explanation, to continue using Hilbert space structures in cognition.

\section*{Acknowledgements}
This work was supported by QUARTZ (Quantum Information Access and Retrieval Theory), the Marie Sk{\l}odowska-Curie Innovative Training Network 721321 of the European Union's Horizon 2020 research and innovation programme.

\end{document}